\documentstyle[aps,aipbook,
psfig]{revtex}
\begin{document}


\title{Analysis of the Systematic Errors in the Positions of
BATSE Catalog Bursts}

\author{C.~Graziani$^1$ and D.~Q.~Lamb$^2$}
\address{
$^1$NASA Goddard Space Flight Center, Greenbelt, MD 20771\\
$^2$Dept. of Astronomy and Astrophysics, University of Chicago,
Chicago, IL 60637\\
}

\maketitle

\begin{abstract}
We analyze the systematic errors in the positions of bursts in the
BATSE 1B, 2B and 3B catalogs, using a likelihood approach.  We use
the BATSE data in conjunction with 196 single IPN arcs.  We assume
circular Gaussian errors, and that the total error is the sum in
quadrature of the systematic error $\sigma_{\rm sys}$ and statistical
error $\sigma_{\rm stat}$, as prescribed by the BATSE catalog.  We find
that the 3B burst positions are inconsistent with the value
$\sigma_{\rm sys} = 1.6^\circ$ stated in the BATSE 3B catalog.
\end{abstract}
\vskip -8pt

\section*{Introduction}
\vskip -8pt
The stated systematic error in the BATSE 3B catalog\cite{Meegan95}
burst locations is $\sigma_{\rm sys} = 1.6^\circ$.  This value was
estimated by taking the RMS deviation of BATSE positions from known
positions of 36 bursts, determined by the IPN, WATCH, and
COMPTEL\cite{Pend95}.  Unfortunately, many of these same known
positions were used to calibrate the BATSE burst location software,
i.e., as a guide in determining what effects to include in the burst
location algorithm.  In order to calibrate properly the BATSE burst
position errors, an independent set of burst locations is needed.  Such
a set exists, in the form of 196 bursts for which single IPN arcs
exist\cite{KH95}.  In this paper we analyze the systematic errors in
the positions of bursts in the BATSE 1B, 2B and 3B catalogs using these
196 bursts.  Our analysis is based upon the likelihood approach.


\vskip -8pt
\section*{Analysis}
\vskip -8pt
We assume that the systematic error $\sigma_{\rm sys}$ and statistical
error $\sigma_{\rm stat}$ in position are circular Gaussians, to be
added in quadrature, as prescribed by the BATSE catalog.  The circular
approximation should be good, since roughly $2/3-3/4$ of all BATSE
bursts have $\chi^2$ positional contours that are nearly
circular\cite{Pend95}, and the bursts with IPN arcs tend to be those
with larger fluences.  We further assume that the sky is flat (i.e.,
that the size of the total error in the BATSE burst position is not too
large), and that the IPN arcs have zero width.  The former is a good
approximation, since the largest BATSE $\sigma_{\rm tot}\approx 12^\circ
\ll 90^\circ$ among the bursts with IPN arcs.  The latter is a good
approximation since the characteristic width of the 3-$\sigma$ contours
of the IPN arcs is a few arcminutes, which is much less than the
smallest BATSE $\sigma_{\rm tot} \approx 1^\circ$.

\begin{figure}[t]
\centerline{\psfig{file=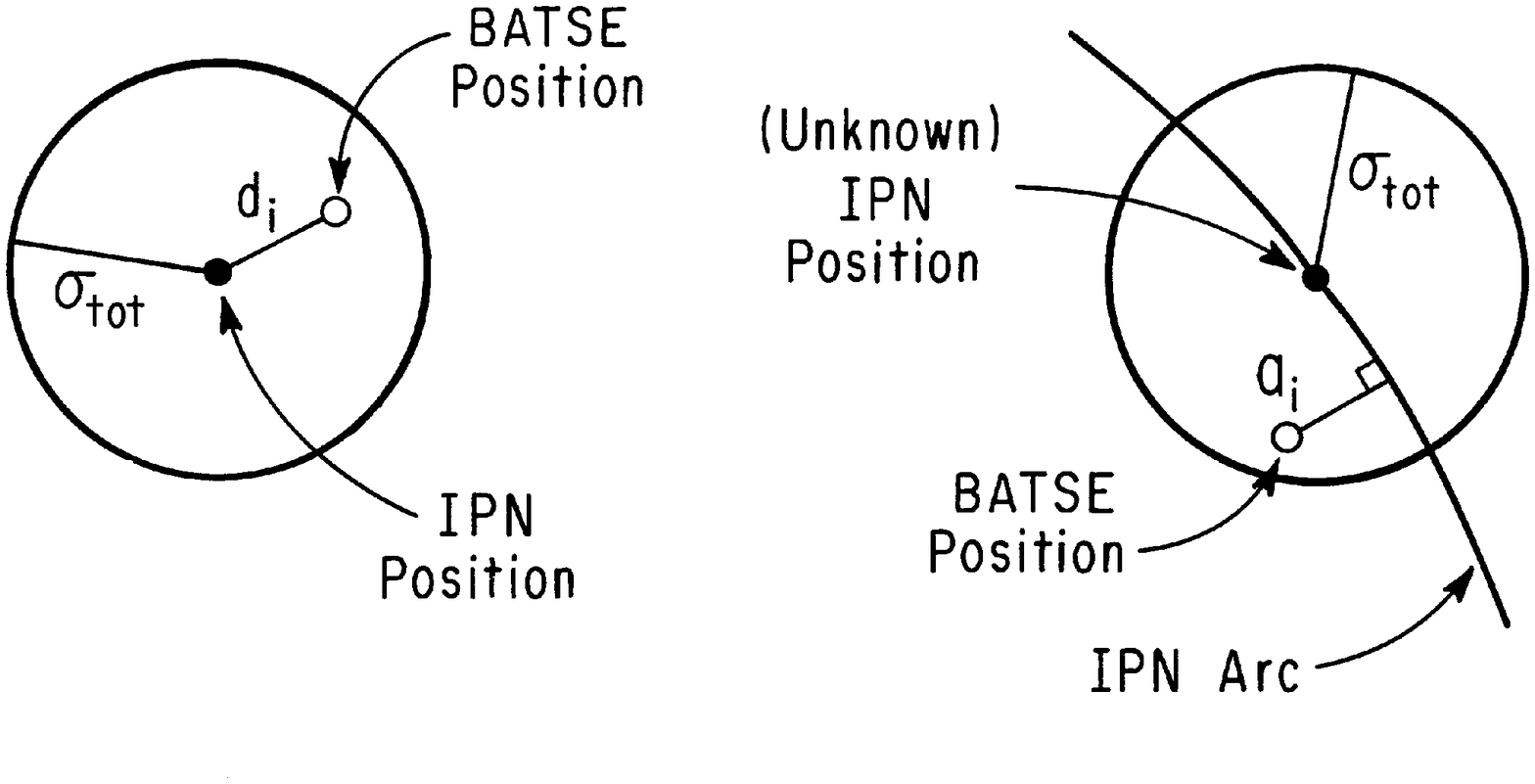,height=1.5in,angle=0} \hfil \psfig{file=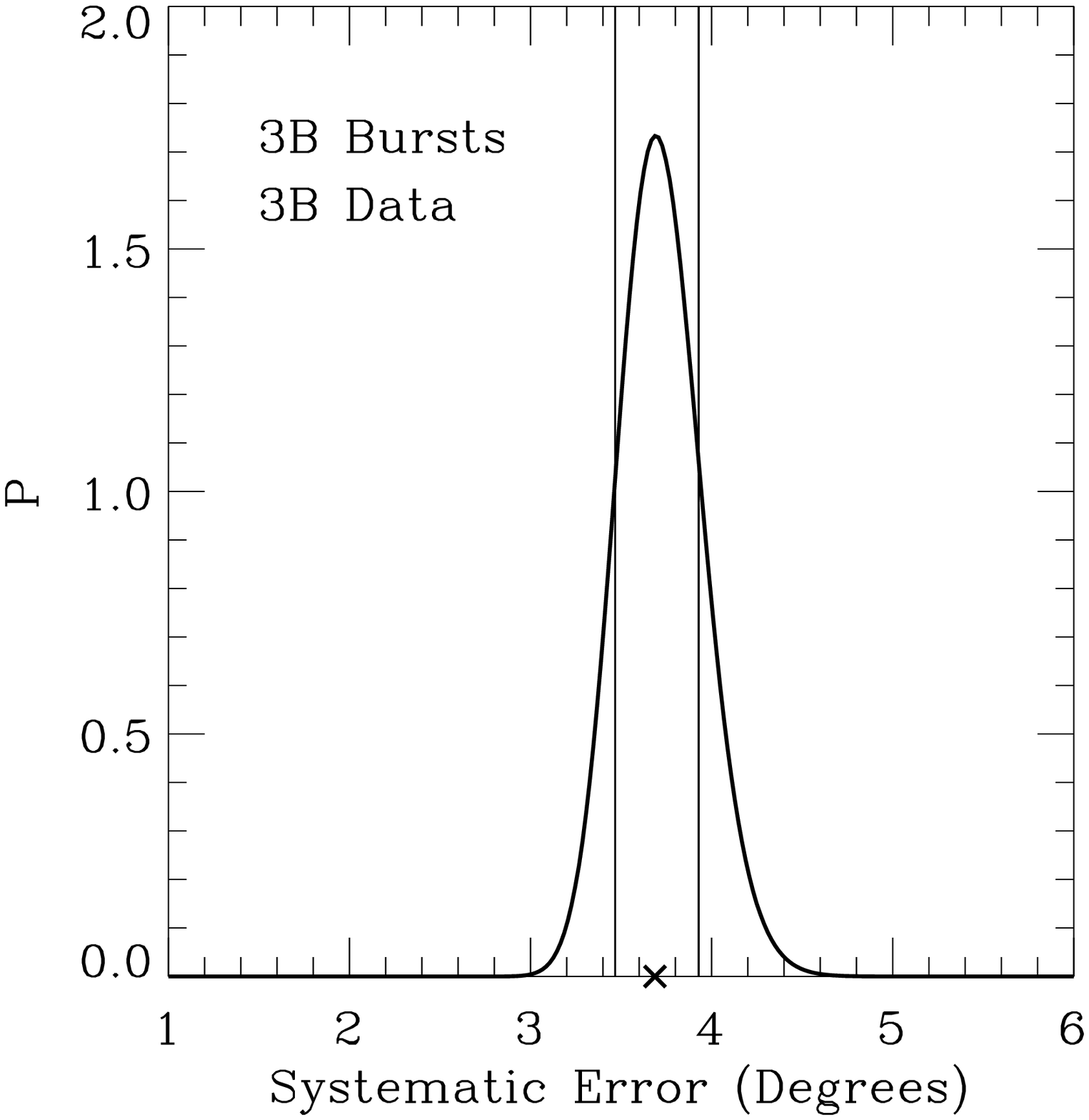,height=1.5in,angle=0}}
{Figure~1.  Configurations for a burst with a known position (left
diagram) and for a burst with a single IPN arc (right diagram).}
\refstepcounter{figure} 
{\label{fig1}}
\vskip 3pt
{Figure~2.  Probability density as a function of $\sigma_{\rm sys}$ for
the ML constant $\sigma_{\rm sys}$ model.  The vertical lines
show the $\pm 1$-$\sigma$ interval.}
\refstepcounter{figure} 
{\label{fig2}}
\end{figure}

The likelihood function for bursts with known positions (e.g., two
intersecting IPN arcs) is given by
\begin{equation}
{\cal L} (\sigma_{\rm sys}) = \prod_{i = 1}^{N} L_i \equiv 
\prod_{i = 1}^{N} {1 \over {2\pi \mu_i^2}} 
\exp\big({-{1 \over 2} {d_i^2 \over \mu_i^2}}\big) \; ,
\end{equation}
whereas the likelihood function for bursts with single IPN arcs is given by
\begin{equation}
{\cal L} (\sigma_{\rm sys}) = \prod_{i = 1}^{N_{\rm arc}} L_i \equiv 
\prod_{i = 1}^{N_{\rm arc}} {1 \over {\sqrt{2\pi} \mu_i}} 
\exp\big({-{1 \over 2} {a_i^2 \over \mu_i^2}}\big) \; .
\end{equation}
Here, $d_i$ is the deviation of the BATSE position from the known
position, $a_i$ is the perpendicular deviation of the BATSE position from
the IPN arc (see Figure 1), and $\mu_i = 0.43 \sigma_{\rm tot} = 0.43
(\sigma_{\rm sys}^2 + \sigma_{\rm stat}^2)^{1/2}$.

The likelihood function ${\cal L} (\sigma_{\rm sys})$ allows
exploration of any model that we may have for $\sigma_{\rm sys}$.  Here
we focus on three models: (1) $\sigma_{\rm sys}$ = a constant, (2)
$\sigma_{\rm sys} = A (S/10^{-5}$ erg cm$^{-2})^\alpha$, and (3)
$\sigma_{\rm sys} = A(\sigma_{\rm stat}/1^\circ)^\alpha$.  The first
model has one free parameter ($\sigma_{\rm sys}$); the second and third
models have two free parameters (scale factor $A$ and power law index
$\alpha$).

\vskip -8pt
\section*{Results}
\vskip -6pt
Fitting the constant $\sigma_{\rm sys}$ model to the 18 BATSE 3B bursts
with IPN positions, we find a maximum likelihood (ML) value
$\sigma_{\rm sys}^{\rm ML} =
{1.85^{\circ}}_{-0.22^\circ}^{+0.28^\circ}$.  This is consistent with
the value of $1.6^\circ$ quoted in the BATSE 3B catalog\cite{Meegan95},
which was found using these bursts and some others.

Fitting the constant $\sigma_{\rm sys}$ model to the 196 BATSE 3B
bursts with IPN arcs, we find a ML value $\sigma_{\rm sys}^{\rm ML} =
{3.7^{\circ}}_{-0.22^\circ}^{+0.24^\circ}$.  This is inconsistent (at
the 10$\sigma$ level!)
\begin{figure}[t]
{\leavevmode\psfig{file=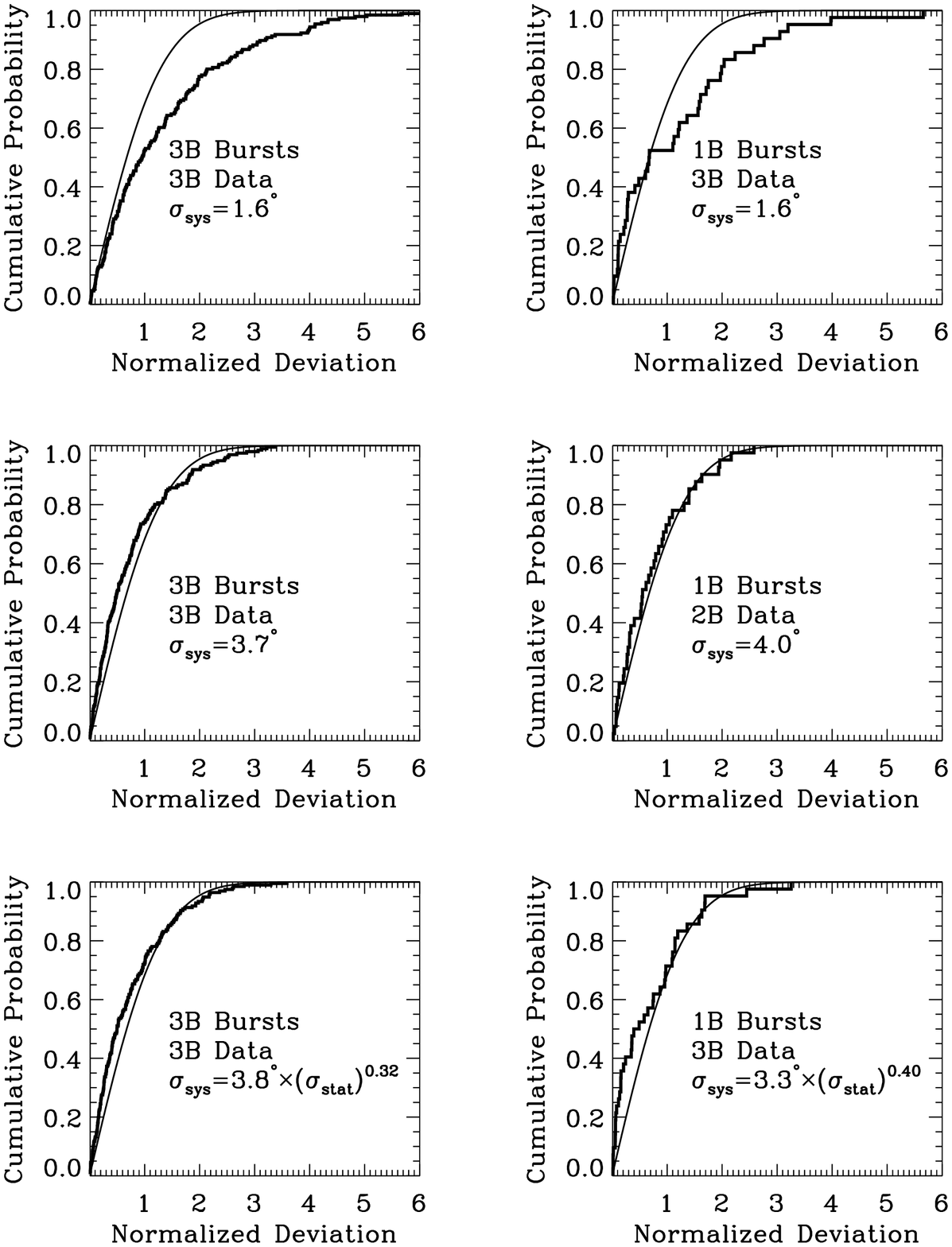,width=4.55in,angle=0}}
{Figure~3.
Comparison of expected and observed cumulative distributions of the
normalized deviation $D = |a_i/\sigma_{\rm tot}^i|$ for the burst
samples and data as labeled, assuming the ML models of
$\sigma_{\rm sys}$ as labeled.
}
\refstepcounter{figure} 
{\label{fig3}}
\end{figure}
\noindent with the value of $1.6^\circ$ quoted in the BATSE 3B catalog,
as shown by the probability distribution in $\sigma_{\rm sys}$ (see
Figure 2) and comparison of the expected and observed distribution of
normalized deviations $D = |a_i/\sigma_{\rm tot}^i|$ assuming
$\sigma_{\rm sys} = 1.6^\circ$ (see Figure 3).

Fitting the second model, in which $\sigma_{\rm sys} = A (S/10^{-5}$ erg
cm$^{-2})^\alpha$, we find
\begin{figure}[m]
{\leavevmode\psfig{file=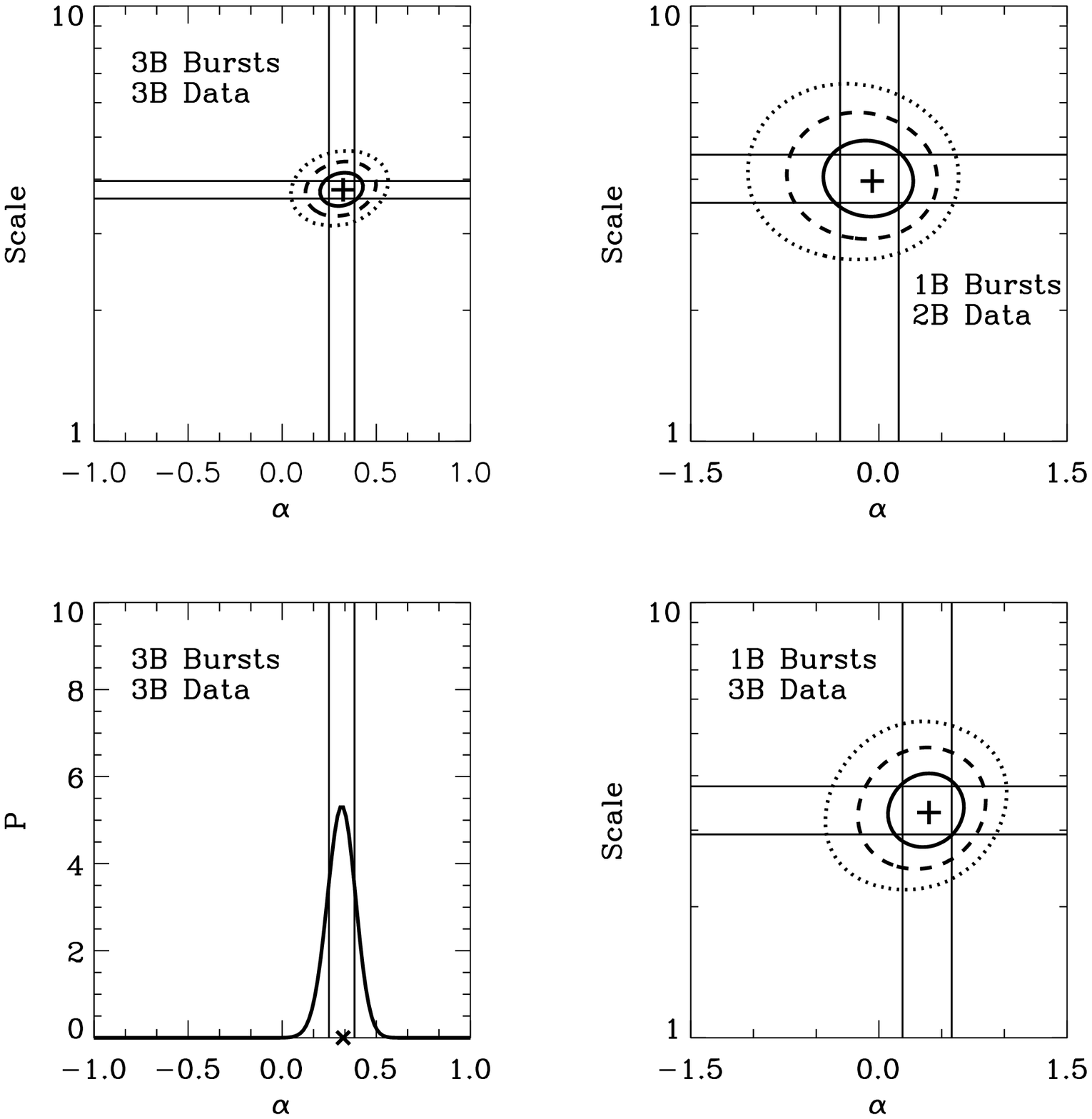,width=4.55in,angle=0}}
{Figure~4.  Upper left-hand and lower and upper right-hand panels:
1-$\sigma$, 2-$\sigma$, and 3-$\sigma$ contours in the
($A,\alpha$)-plane for the burst samples and data as labeled, assuming
the $\sigma_{\rm stat}$-dependent model.  Lower left-hand
panel:  same as Figure 2, except for the $\sigma_{\rm
stat}$-dependent model.
}
\refstepcounter{figure} 
{\label{fig4}}
\end{figure}
\noindent ML values $A_{\rm ML} = 3.8 \pm
0.3$ and $\alpha_{\rm ML} = -0.18_{-0.05}^{+0.06}$.  The probability
$P(\alpha \ge 0) = 1.6 \times 10^{-3}$, showing that the data prefer
the power-law model over the ML model with constant $\sigma_{\rm sys} =
3.7^\circ$.

Fitting the third model, in which $\sigma_{\rm sys} = A \sigma_{\rm
stat}^\alpha$, we find ML values $A_{\rm ML} = 3.8 \pm 0.2$ and
$\alpha_{\rm ML} = 0.32_{-0.08}^{+0.06}$ (see Figure 4).  The
probability $P(\alpha \le 0) = 4.7 \times 10^{-5}$, showing that the
data strongly prefer the power-law model over the ML model with
constant $\sigma_{\rm sys} = 3.7^\circ$.
 
These results imply that $\sigma_{\rm sys}$ is strongly correlated with
$S$ and $\sigma_{\rm stat}$.  For example, $\sigma_{\rm sys}
(\sigma_{\rm stat} = 0.1^\circ) = 2^\circ$ while $\sigma_{\rm sys}
(\sigma_{\rm stat} = 10^\circ) = 8^\circ$.  This is illustrated in
Figure 5, which shows the ML $S$-dependent and $\sigma_{\rm
stat}$-dependent models.

\begin{figure}[t]
{\leavevmode\psfig{file=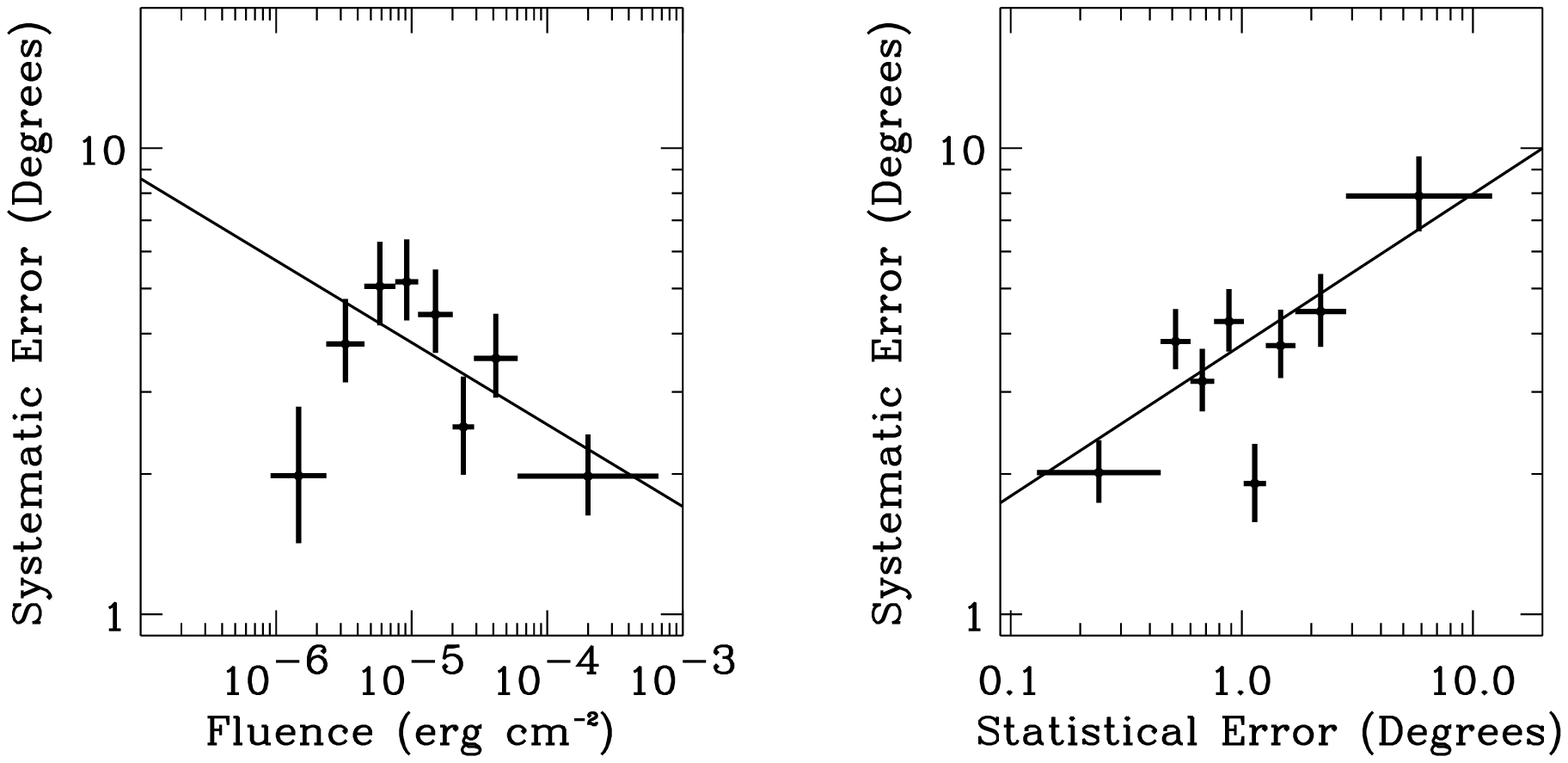,width=4.55in,angle=0}
}
{Figure~5.
Left-hand panel: The ML $S$-dependent model.
Right-hand panel: The ML $\sigma_{\rm stat}$-dependent model.
The points are {\it not} data points, but rather are the ML {\it
constant} $\sigma_{\rm sys}$ model for the given interval in $S$ or
$\sigma_{\rm stat}$.
}
\refstepcounter{figure} 
{\label{fig5}}
\end{figure}

It is also interesting to examine the character of $\sigma_{\rm sys}$
for subsets of the 3B catalog, in particular the 1B sample of bursts in
which evidence for repeating was found.  Fitting the
three models to the ``old'' 1B positions, we find
$\sigma_{\rm sys} = 4.0^\circ$ for the ML constant $\sigma_{\rm sys}$
model, and that the data do not request a more
complicated model(see Figures 3 and 4).  In contrast, fitting the 3B positions
for the 1B sample of bursts, we find that the data request the
$\sigma_{\rm stat}$-dependent model at the 95\% confidence level [e.g.,
$P(\alpha \le 0) = 1.6 \times 10^{-2}$] (again see Figures 3 and 4).

These results imply that for the 1B sample of bursts, the new (3B)
$\sigma_{\rm sys}$'s are smaller than the old (1B) ones for bursts with
$S > 4\times 10^{-6}$~erg~cm$^{-2}$ or with $\sigma_{\rm stat} <
1.6^\circ$.  However, for the $\approx$ 80\% of bursts with lower
fluences or larger $\sigma_{\rm stat}$'s, they imply that the new
$\sigma_{\rm sys}$'s are larger than the old ones.

\medskip

We are grateful to Kevin Hurley for providing us with the 3rd catalog
of IPN arcs, without which this analysis would not have been possible.
We also wish to thank Geoff Pendleton and Michael Briggs for discussing
BATSE burst location issues with us.

\end{document}